# Structural, magnetic and electronic properties of quaternary oxybismuthides LaOMBi (where *M* = Sc, Ti ... Ni, Cu) - possible parent phases for new superconducting materials

I.R Shein[*], V.L. Kozhevnikov and A.L. Ivanovskii

*Institute for Solid State Chemistry, Ural Branch of the Russian Academy of Sciences, Ekaterinburg 620041, GSP-145, Russia*

(Dated: April 25, 2008)

**The extensive *ab initio* total energy calculations using the VASP-PAW method with the generalized gradient approximation (GGA) for the exchange-correlation potential are applied to systematic investigation of structural, electronic and magnetic properties in quaternary oxybismuthides LaOMBi (where *M* = Sc, Ti…Ni, Cu). The energy spectrum features similar to lanthanum-iron oxyarsenide LaOFeAs and non-magnetic ground state are indicative of superconductivity possible in lanthanum-nickel oxybismuthide LaONiBi.**

New layered superconductor, fluorine doped lanthanum-iron oxyarsenide LaO$_{1-x}$F$_x$FeAs with $T_C$ of about 26K, was reported in February 2008 [1]. This finding has attracted a great deal of interest because it seems to have set a first example of long sought for copper-free compounds with properties combination (high critical temperature, large upper critical field, bordering magnetic instability) that suggest unconventional superconductivity [2-7]. Also it has triggered much activity in search of related materials. As doping and chemical substitutions are known to be powerful tools for properties tailoring, it is not surprising that soon after these methods have resulted in discovery of strontium doped derivatives La$_{1-x}$Sr$_x$OFeAs [8]. Further substitution of lanthanum for other rare earths brought forth GdO$_{1-x}$F$_x$FeAs [9], CeO$_{1-x}$F$_x$FeAs [10], SmO$_{1-x}$F$_x$FeAs [11] with critical temperature as high as 50-52K in PrO$_{1-x}$F$_x$FeAs and NdO$_{1-x}$F$_x$FeAs [12,13]. A set of superconductors was discovered also very recently for related fluorine-free systems with oxygen non-stoichiometry with $T_C$ up to 53K [14,15].

E-mail address: shein@ihim.uran.ru

LaOFeAs adopts a layered tetragonal crystal structure (ZrCuSiAs type, space group P4/*nmm*, Z = 2 [16]), where layers of (La-O) are sandwiched with (Fe-As) layers. According to DFT calculations [5,17-19] for LaOFeAs the electronic bands in a window around the Fermi level are formed mainly by the states of (Fe-As) layers while the bands of (La-O) layers are rather far from the Fermi level. Thus, variations in inter- and intra-layer distances and in chemical composition all seem to be important factors for optimizations of properties in these materials.

Some experimental [20] and theoretical [11,22] attempts have been made already for understanding composition dependent behavior of oxyarsenides *Ln*OFeAs (*Ln* =La, Ce…Tm) [20,21] and LaO*M*As (*M* = V, Cr…Ni, Cu) [22].

In this Communication we report on results of systematic studies in structural, electronic and magnetic properties for tetragonal quaternary oxybismuthides LaO*M*Bi (where *M* = Sc, Ti…Ni, Cu) as possible parent phases for materials similar to superconducting oxyarsenides.

Our calculations were performed within the projector augmented wave (PAW) method [23] in formalism of density functional theory as implemented in the VASP package [24-26]. We used the exchange-correlation functional with the generalized gradient approximation (GGA) in PBE form [27]. The calculations were performed with full lattice optimizations including internal coordinates, $z_{La}$ and $z_{Bi}$, which define inter-layer distances La-O and *M*-Bi, respectively [16]. The calculations were performed for nonmagnetic (NM) and ferromagnetic (FM) states.

The optimized lattice parameters and internal coordinates for LaO*M*Bi are shown in Table 1. As compared with LaO*M*As [22], the lattice constants for the corresponding oxybismuthides are about 2-7 % larger, which is a clear manifestation of the difference between bismuth (1.82 Å) and arsenic (1.48 Å) atomic radii.

The calculated stabilization energies (ΔE) of FM state relative to NM state, $\Delta E_{(FM-NM)} = E^{tot}_{FM} - E^{tot}_{NM}$, are shown in Figure 1. Two different groups of oxybismuthides can be clearly identified. The non-magnetic phases LaOMBi, where M = Sc, Ti, Ni and Cu, and all others where spin polarization will be favored and magnetic ground states are stabilized.

Figure 1 shows also variations in magnetic moments (MMs) for 3$d$ metals in LaOMBi compared with those in LaOMAs [22]. One can see the features as follows:

i. These dependencies are nonlinear with maximal MMs in Mn-containing phases;

ii. The MM values are larger for oxybismuthides than for corresponding oxyarsenides;

iii. The calculations predict strong magnetic behavior for Fe-containing oxybismuthide (with magnetic moment of 0.84 $\mu_B$ on iron) while NM or weak anti-ferromagnetic states is expected to be stable in LaOFeAs [5,17-19,22].

Let's now discuss the above mentioned tendencies with the using of the picture of electronic structure in oxybismuthides. It is seen in Figures 2 and 3 that the valence DOSs for LaOMBi can be divided into three main regions: (i) from -5.5 eV to -2.8 eV with mainly oxygen 2$p$ states, (ii) from -2.8 eV to -0.5 eV with strongly hybridized M 3$d$ and Bi states, and finally (iii) from -0.5 eV to $E_F$ = 0 eV with mainly M 3$d$ states.

A monotonous increase of MMs in magnetic LaOMAs can be seen when moving from V to Mn. This change is related to predominant filling of the spin-up states, Figure 3, while majority of spin-down states remain partially empty. The largest MM is observed for LaOMnBi. At the end of the 3$d$ row (Fe and Co) the 3$d$ spin-down bands begin to fill, and this leads to a monotonous decrease in MM values so that spin splitting completely disappears in LaONiBi.

Thus, electronic and magnetic properties in oxybismuthides (LaO$M$Bi) and oxyarsenides (LaO$M$As [22]) seem quite similar to each other in many respects. At the same time replacement of arsenic for bismuth results in a larger cell volume that, in turn, leads to some changes for 3$d$ band splitting and in increase of atomic magnetic moments for LaO*M*Bi. Most significant, qualitative difference is seen for the case of iron when LaOFeAs does not bear a magnetic moment while LaOFeBi seems to have appreciable magnetic moment on iron.

Thus, while in oxyarsenides family iron containing LaOFeAs occurs at the border of magnetic instability, the similar state for oxybismuthides ought to develop in nickel containing LaONiBi. The calculated total DOS values at the Fermi level $N(E_F)$, Sommerfeld coefficients $\gamma = (\pi^2/3)N(E_F)k_B^2$ and molar Pauli paramagnetic susceptibility $\chi = \mu_B^2 N(E_F)$ within free electron model for oxybismuthides are shown in Table 2. The obtained results clearly demonstrate that these parameters are quite similar in LaONiBi and LaOFeAs [22]. Therefore, a preliminary conclusion can be drawn that oxybismuthide LaONiBi may possibly be considered as favorable matrix for a superconducting material.

Finally, the present Communication deals with ideal oxybismuthides. It is clear that further in-depth studies are necessary in order to make more certain predictions. For instance, the influence of both donor and acceptor type doping on electronic and magnetic properties is of great interest. Also, coexistence of ferromagnetic and antiferromagnetic states in oxybismuthides should be considered. The last point is important, in particular, in view of proposed scenario for unconventional electron pairing mechanism (mediated by antiferromagnetic spin fluctuations [5]) in oxypnictides and also for systematic exploration of relationships between magnetism and superconductivity.

**Table 1.** Optimized lattice parameters (a and c, Å), internal coordinates ($z_{La}$ and $z_{Bi}$), stabilization energies for ferromagnetic states in comparison with nonmagnetic states ($\Delta E_{(FM-NM)} = E^{tot}_{FM} - E^{tot}_{NM}$, eV/cell) and magnetic moments for the 3$d$ atoms (MM, $\mu_B$) for quaternary oxybismuthides LaO$M$Bi, where $M$ = Sc, Ti, V,…Ni, Cu.

| d-metal / parameter | $a$ | $c$ | MM |
|---|---|---|---|
| Sc | 4.3847 | 9.6891 | 0 |
| Ti | 4.2562 | 9.7756 | 0 |
| V  | 4.0919 | 10.0990 | 0.437 |
| Cr | 4.2662 | 9.4083 | 2.943 |
| Mn | 4.2106 | 9.8361 | 3.640 |
| Fe | 4.1741 | 9.2029 | 0.844 |
| Co | 4.2306 | 8.9522 | 0.499 |
| Ni | 4.2616 | 8.8878 | 0 |
| Cu | 4.0871 | 10.2975 | 0 |
| d-metal / parameter | $z_{La}$ | $z_{Bi}$ | $-\Delta E_{(FM-NM)}$ |
| Sc | 0.1100 | 0.7023 | 0 |
| Ti | 0.1164 | 0.6912 | 0 |
| V  | 0.1204 | 0.6853 | 0.013 |
| Cr | 0.1194 | 0.6811 | 0.322 |
| Mn | 0.1168 | 0.6840 | 0.702 |
| Fe | 0.1269 | 0.6667 | 0.031 |
| Co | 0.1284 | 0.6623 | 0.024 |
| Ni | 0.1278 | 0.6640 | 0 |
| Cu | 0.1175 | 0.6744 | 0 |

**Table 2.** Total DOSs at the Fermi level ($N(E_F)$, in states/eV·mol.unit), Sommerfeld coefficients ($\gamma$, in mJ/(mol·K$^2$)) and molar Pauli paramagnetic susceptibility values ($\chi$, in $10^{-4}$ emu/mol) in oxybismuthides LaOMBi, where $M$ = Sc, Ti,…Ni, Cu.

| d-metal / parameter | $N(E_F)$ | $\gamma$ | $\chi$ |
|---|---|---|---|
| Sc | 4.72 | 11.13 | 1.53 |
| Ti | 3.31 | 7.80 | 1.07 |
| V * | 1.91 | 4.50 | 0.62 |
| Cr * | 2.34 | 5.52 | 0.76 |
| Mn * | 2.22 | 5.24 | 0.72 |
| Fe * | 2.73 | 6.44 | 0.88 |
| Co | 1.69 | 3.98 | 0.55 |
| Ni | 2.69 | 6.34 | 0.87 |
| Cu | 1.02 | 2.40 | 0.33 |

- for FM states.

**FIGURES**

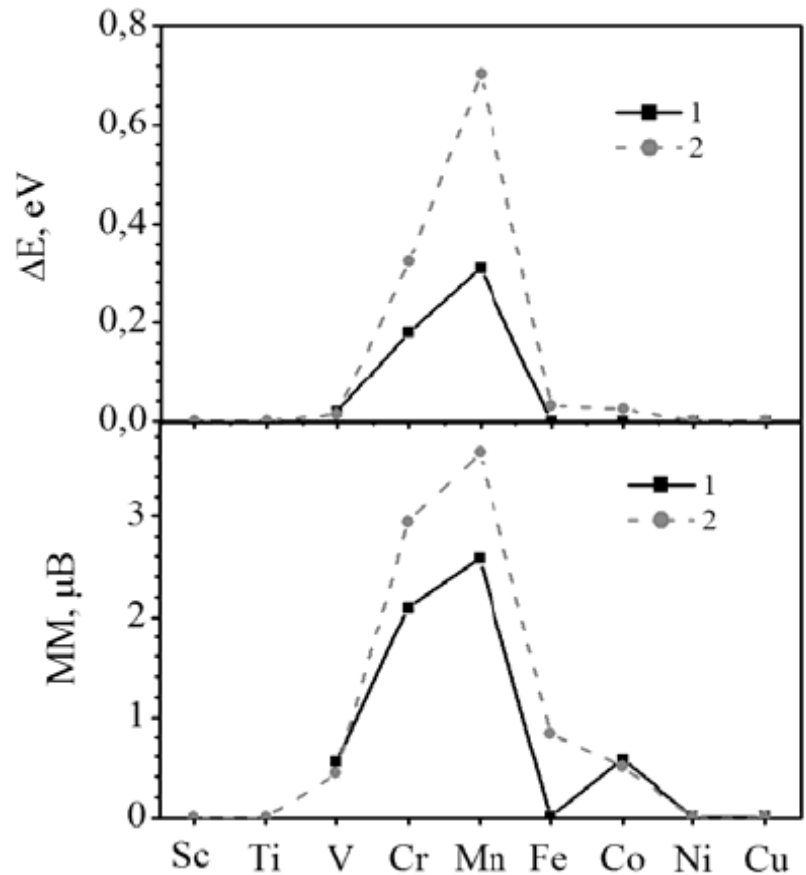

**Figure 1.** The calculated magnetic moments for 3$d$ atoms and stabilization energies for ferromagnetic states in comparison with nonmagnetic states in quaternary oxybismuthides LaO$M$Bi, where $M$ = Sc, Ti, V,…Ni, Cu (2, dotted lines) in comparison with quaternary oxyarsenides LaO$M$As [22] (1, solid lines).

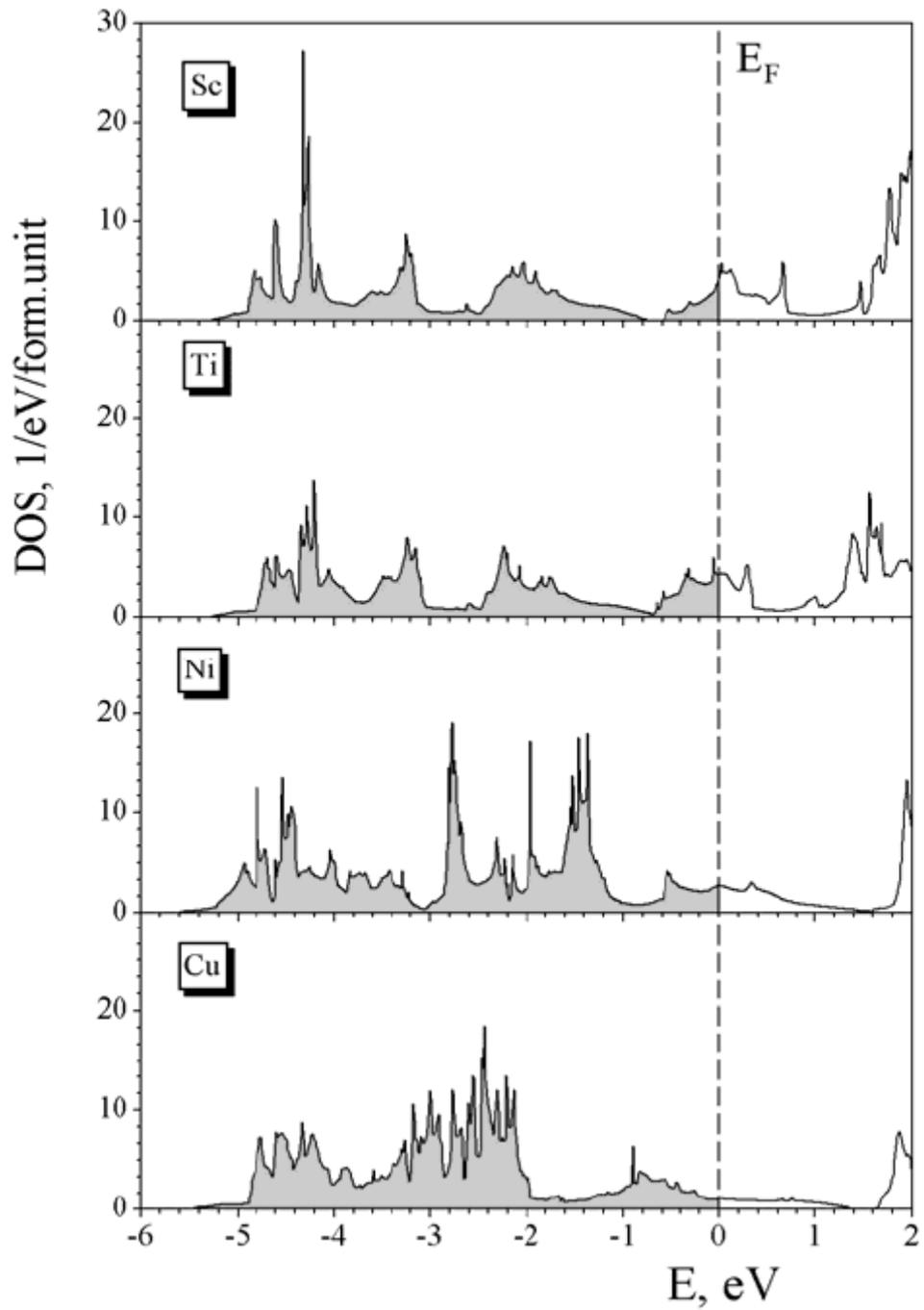

**Figure 2.** The calculated electron density of states (DOS) for nonmagnetic oxybismuthides LaO$M$Bi, where $M$ = Sc, Ti, Ni, and Cu.

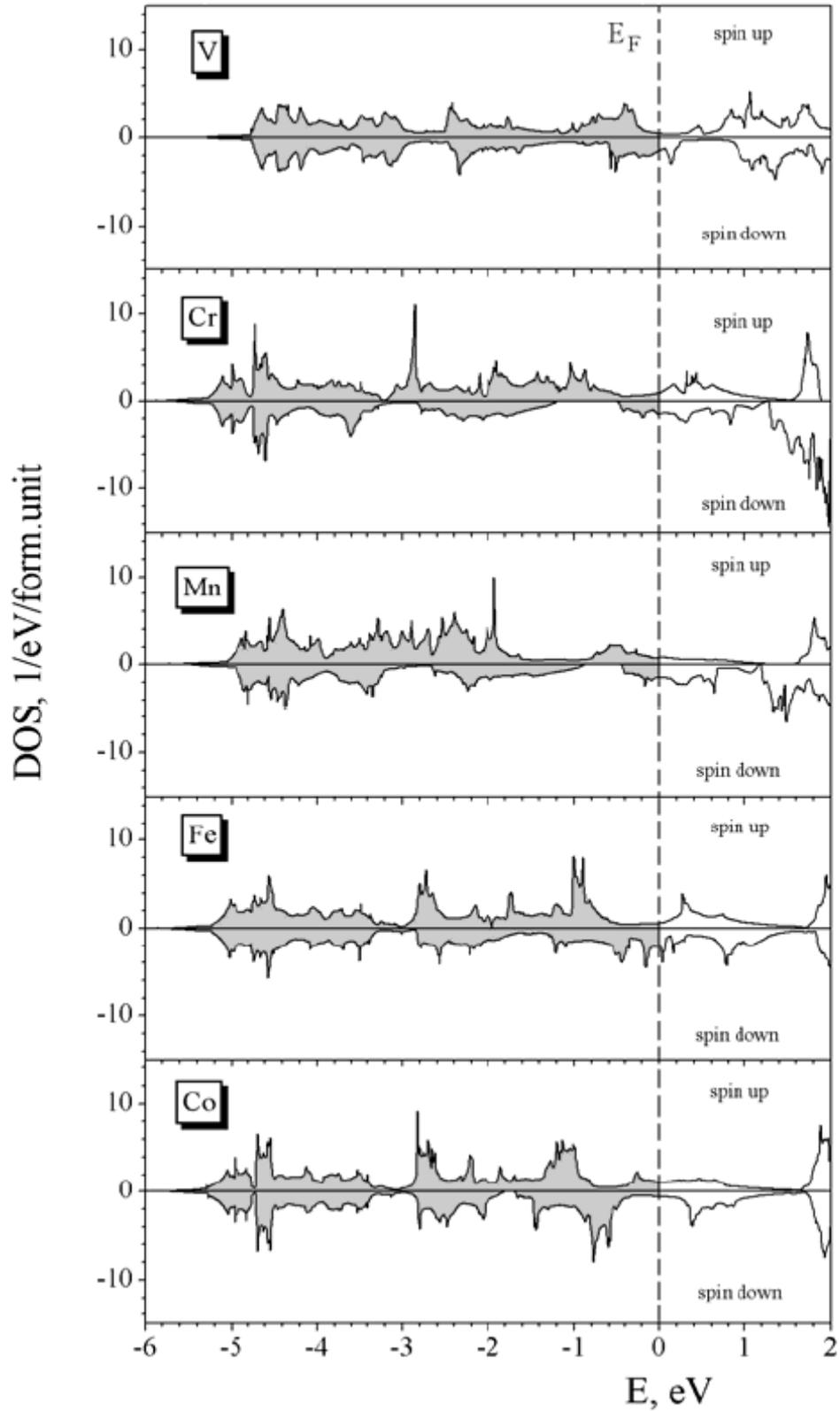

**Figure 3.** The calculated electron density of states (DOS) for magnetic oxybismuthides LaO$M$Bi, where $M$ = V, Cr, Mn, Fe and Co. (FM state)